\documentclass[11pt]{article}
\usepackage{amsmath,amssymb,color}
\usepackage{amsfonts}
\newcommand{\hoch}[1]{$\, ^{#1}$}

\makeatletter
\@addtoreset{equation}{section}
\makeatother

\newcommand{\be}{\begin{equation}}
\newcommand{\ee}{\end{equation}}
\newcommand{\bea}{\setlength\arraycolsep{2pt} \begin{eqnarray}}
\newcommand{\eea}{\end{eqnarray}}
\newcommand{\nn}{\nonumber}

\def\crampest{\medmuskip = 1mu plus 1mu minus 1mu}
\def\uncramp{\medmuskip = 4mu plus 2mu minus 4mu}

\def\ft#1#2{{\textstyle{\frac{\scriptstyle #1}{\scriptstyle #2} } }}
\def\fft#1#2{{\frac{#1}{#2}}}

\def\0{{\sst{(0)}}}
\def\1{{\sst{(1)}}}\def\2{{\sst{(2)}}}
\def\3{{\sst{(3)}}}
\def\4{{\sst{(4)}}}
\def\5{{\sst{(5)}}}
\def\6{{\sst{(6)}}}
\def\7{{\sst{(7)}}}
\def\8{{\sst{(8)}}}
\def\sst#1{{\scriptscriptstyle #1}}

\def\del{{\partial}}
\def\im{{{\rm i}}}

\begin{document}

\begin{flushright}
\hfill UPR-1252-T\ \ \ \ MIFPA-13-19
\end{flushright}

\vspace{25pt}
\begin{center}
{\Large {\bf Entropy-Product Rules for Charged Rotating Black Holes}}

\vspace{20pt}
{\Large M. Cveti\v c\hoch{1,2}, H. L\"u\hoch{3} and C.N. Pope\hoch{4,5}}

\vspace{10pt}

\hoch{1}{\it Department of Physics and Astronomy,\\
 University of Pennsylvanian, Philadelphia, PA 19104, USA}

\vspace{10pt}

\hoch{2}{\it Center for Applied Mathematics and Theoretical Physics,\\
University of Maribor, Maribor, Slovenia}

\vspace{10pt}

\hoch{3}{\it Department of Physics, Beijing Normal University,
Beijing 100875, China}

\vspace{10pt}

\hoch{4}{\it George and Cynthia Woods Mitchell Institute for Fundamental
Physics and Astronomy\\
Texas A\&M University, College Station, TX 77843-4242, USA}

\vspace{10pt}

\hoch{5}{\it DAMTP, Centre for Mathematical Sciences,
 Cambridge University,\\  Wilberforce Road, Cambridge CB3 OWA, UK}

\vspace{40pt}

\underline{ABSTRACT}
\end{center}

We study the universal nature of the product of the entropies of all 
horizons of 
charged rotating black holes.  We argue, by examining further explicit 
examples, that when the maximum number of rotations and/or charges are 
turned on, the entropy product is expressed in terms of angular momentum 
and/or charges only, which are quantized. (In the case of gauged 
supergravities, the entropy product depends on the gauge-coupling constant
also.)  In two-derivative gravities, the notion of the ``maximum number'' of 
charges can be defined as being sufficiently many non-zero charges that
the Reissner-Nordstr\" om black hole arises under an appropriate 
specialisation of the charges. (The definition can be relaxed somewhat 
in charged AdS black holes in $D\ge 6$.)  In higher-derivative gravity, we 
use the charged rotating black hole in Weyl-Maxwell gravity as an 
example for which the entropy product is still quantized, but it is expressed  
in terms of the angular momentum only, with no dependence on the charge.  
This suggests that the notion of maximum charges in 
higher-derivative gravities requires further understanding.

\thispagestyle{empty}

\pagebreak

\tableofcontents
\addtocontents{toc}{\protect\setcounter{tocdepth}{2}}

%%%%%%%%%%%%%%%%%%%%%%%%%%%%%%%%%%%%%%%%

%\newpage
%%%%%%%%%%%%%%%%%%%%%%%%%%%%%%%%%%%%%%%%

\section{Introduction}

  Understanding black hole entropy at the microscopic level has been a major
focus of research in string theory and M-theory in the past years. While the  
microscopics of asymptotically-flat BPS black holes in four and five 
dimensions is by now well understood \cite{SV}  (for a review, see, for 
example, \cite{Sen} and references therein),  the internal properties of 
general non-extremal black holes are less clear.  However, it has been 
known for a long time that  general asymptotically-flat multi-charged 
rotating black holes of supergravities in four \cite{CYII}  and 
five \cite{CYI} dimensions\footnote{These black holes can be used as
generating solutions for the maximally supersymmetric
${\cal N}=4$ (${\cal N}=8$) supergravities obtained by toroidally
compactifying the heterotic string (or Type IIA string or M-theory).
In addition to the mass $M$,  these solutions are specified in four
dimensions by four charges $Q_i$ ($i = 1,2,3,4$) and one angular
momentum $J$.  In five dimensions they are specified by the mass and
three charges, $Q_i$ ($i = 1, 2, 3$),
and two angular momenta, $J_1$ and $J_2$. It turns out that in four
dimensions the complete generating solution is specified by an additional
fifth charge, which has been obtained only in the BPS \cite{CT}  and
static  \cite{CYIII} cases.}  have a  tantalizing entropy formula
\cite{CYII},  and a first law of thermodynamics \cite{L,CLI,CLII}
associated with both the inner and the outer black hole horizons,
which are  highly  suggestive of a possible microscopic interpretation in
terms of a two-dimensional conformal field theory (CFT).  Specifically,
the entropies $S_{\pm}$  of the outer and inner horizons are  of the
form \cite{CYII,CLI,CLII}: $S_\pm=2\pi(\sqrt{N_L}\pm\sqrt{N_R})$, where
the quantities  $N_L$ and $N_R$ may be viewed as the excitation numbers
of the left and right moving modes of a weakly-coupled two-dimensional
conformal field theory.  The product  $S_+\, S_- = 4\pi^2(N_L-N_R)$  should
therefore by quantised  in
integer multiples of $4\pi^2$  \cite{L, CLI,CLII} (and
re-emphasized  in  \cite{CLIII}).   Indeed, one finds:
%%%%%
\bea
S_+\, S_- &= &4\pi^2\Big(J^2+ \prod_{i=1}^4 Q_i \Big)\,, \label{prod4}\\
S_+\, S_- &= &4\pi^2\Big(J_1 J_2 + \prod_{i=1}^3 Q_i\Big)\,,
\eea
%%%%%
for  four and five dimensional black holes, respectively.  (These results
were implicit in
\cite{CLI,CLII},  although not explicitly  evaluated.) These expressions are
modulus-independent,  and are   expressed solely in terms of the quantized
duality-invariant quartic (cubic) charge form, and the
quantised angular momenta.

   In  parallel developments  Ansorg and collaborators (see, for
example, \cite{Ansorg1,Ansorg7}  and references therein)  studied
axisymmetric solutions  of Einstein-Maxwell gravity, with sources external
to the outer horizon. They  obtained  striking universal formulae
expressing the entropy products  of the outer  and inner  Killing horizons
in terms of  the total angular momentum $J$ and total charge $Q$. For the
Reissner-Nordstr\"om black hole \cite{Ansorg1},  these products reduce
to (\ref{prod4}) with all $Q_i=Q$. The quantized nature of the 
entropy-product  formulae for other asymptotically-flat solutions, such 
as  general ring and  string solutions, was recently verified in \cite{CR}, and for  static black holes and rings  of  N=2 supergravity in four and five dimensions in \cite{OrtinI} and \cite{OrtinII, OrtinIII}, respectively.

     Another  approach  that has brought considerable insights  into
the internal structure of general black holes is the study of
absorption coefficients or greybody factors for fields in the black hole
background.  This involves solving the wave equations for external fields
in the black hole geometry.
A remarkable feature of many black-hole metrics is
that the wave equations, such as the Klein-Gordon equation for a
minimally coupled massless scalar
field, are typically separable, and this greatly simplifies the study of the
scattering problem.  The core of the calculation is reduced to the
investigation of the solutions of the radial equation, whose complexity is
governed by the nature of its singular points.
As well as having singular
points at the origin and at infinity, additional singularities occur
at all of the zeros of the metric functions that determine the number, and the
locations, of the horizons.  Thus it can be that important features of the
scattering process are governed not only by the properties of the metric
outside and on the outer horizon, but also by its properties at interior
horizons and at other singular points of the metric radial
functions.

Specifically,  the radial part of Klein-Gordon equation for massless probe
scalars in  the background of general asymptotically-flat black holes
in four and five dimensions exhibits an approximate $SL(2,{\bf R})
\times SL(2,{\bf R})$ conformal symmetry, associated with the poles at the
inner and outer black hole horizons \cite{CLI, CLII,CMS}.  The terms that
break this symmetry  are associated with features of the asymptotic geometry,
and can be neglected in an appropriate low-energy regime for
the probe scalars.  This raises the expectation \cite{CMS} that at least
the low-energy dynamics of general black holes could be described by a
two-dimensional CFT.\footnote{ These terms can  also be neglected for
special black-hole backgrounds, including the near-supersymmetric limit
(the AdS/CFT correspondence) \cite{MS, CLI,CLII}  and the near-extreme
rotating limit (the Kerr/CFT correspondence) \cite{GHSS,CLIII}.}

    Recently, this proposal was developed further in \cite{CL11I,CL11II} ,
by identifying an explicit part of the general multi-charged rotating
black-hole geometry that exhibits a manifest  $SL(2,{\bf R})\times
SL(2,{\bf R})$ conformal symmetry of the wave equation. The metrics of
these conformal backgrounds differ  from the original black-hole metrics
by the removal of certain terms in the warp factor only,  and they were
accordingly
dubbed the ``subtracted geometries.''\footnote{The sources for the
subtracted geometry were obtained in \cite{CG} as a certain scaling limit
of another black hole. The full solution for the subtracted geometry
can also be obtained by acting on the original black-hole solution with
specific Harrison transformations \cite{CG,Virmani,CGZ,VirmaniII}
within the STU model, which is a consistent truncation of maximally
supersymmetric supergravity.  For related works, see \cite{KI,KII,KIII}.}
The key global structure and the  thermodynamic properties of these
subtracted geometries, such as the areas of the two horizons and
the angular periodicities, remain the same, and so the subtracted
geometry is expected to preserve  the information about 
internal structure of the black hole.
The subtracted geometry is, however, 
asymptotically conical \cite{CL11II,CG}, rather than asymptotically flat.
A physical interpretation of  the subtraction is
the removal of the  ambient asymptotically Minkowski space-time in a
way that extracts the ``intrinsic''  $SL(2,{\bf R})\times SL(2,{\bf R})$
symmetry of the black hole.
 A lift  of the subtracted metric on a circle gives rise to
AdS$_3\times$ Sphere geometries, and thus the microscopic interpretation 
of the general black-hole entropy can be deduced via an AdS$_3$/CFT$_2$
correspondence \cite{CL11I,CL11II}.
Further studies of the  properties of the dual CFT operators
that parameterise deformations  from the subtracted geometry  were
carried out in \cite{deBoer,CGZ}.

    The intriguing  internal properties of general asymptotically-flat
black holes in four and five dimensions, and their potential
dual two-dimensional CFT descriptions, are intimately related to
geometrical properties of the two horizons.  On the other hand,  black
holes in asymptotically anti-de Sitter (AdS) spacetime, and  rotating
black holes in dimensions larger then five,   have the property that
the radial metric function have more than two zeros.\footnote{To be more
precise, the singular points we are referring
to correspond to all the (real or complex)
values $r_i$ of the radial coordinate $r$
at which the norm of some Killing
vector vanishes.  The metric on the surface at the fixed radius $r_i$
may have signature $(0,+,+,+,\ldots,+)$, in which case it is an ordinary
horizon; or signature $(0,-,+,+,\ldots, +)$, in which case the surface is
a timelike ``pseudo-horizon'' with imaginary area; or $r_i$ may be complex
(such roots arise in conjugate pairs).
For the sake of brevity, in this paper we shall refer to all of the
surfaces defined by the roots of the relevant metric radial function as
``horizons.''}
The wave equations in these backgrounds
will have dominant contributions associated with poles at each of these
zeros. One can therefore again expect that the thermodynamics associated
with {\it each} pole will play a role in governing the properties of the
black hole at the microscopic level. This  could potentially be
suggestive of a microscopic behaviour of such black holes in terms
of a dual field theory in more than two dimensions.
One specific (mesoscopic)  test of these ideas is the calculation of the
product of {\it all} the horizon entropies \cite{cvgipo}.  It turns out
that these entropy products are also universal; they
 depend only on quantized charges, quantized angular momenta and
the cosmological (or gauge-coupling)  constant, which is also quantized
in the context of compactifications of string theory.

   Most of the black-hole examples that have been investigated arise as
solutions of conventional gravity or supergravities with second-order
equations of motion, for which the horizon area and entropy are related by
the Bekenstein-Hawking formula $S=\ft14 A$.  In these cases the quantisation
of the product of entropies is therefore synonymous with the quantisation of
the product of horizon areas.  In higher-derivative gravities, by contrast,
the entropy is no longer in general proportional to the area of the
horizon, but is instead given by the Wald formula which involves the
variation of the action with respect to the Riemann tensor.
Entropy-product formulae in higher-derivative
gravities were studied recently in \cite{CDGK}.
A question arises as to whether in higher-derivative theories
it is the product of the entropies or the product of the areas (or neither)
that is quantised.
Subtleties can arise when trying to answer this question.
In particular, the definition of entropy can be ambiguous
in any even spacetime dimension, since one can always
add a purely topological Euler
integrand
to the action which, while not affecting the equations of motion,
does change the Wald entropy by a
purely numerical additive constant.  For example, in four dimensions
one can add
a Gauss-Bonnet term  to the standard Einstein-Hilbert plus matter
Lagrangian, such
that the original entropy $S_0^i$ at the $i$'th horizon is modified to
%%%%%
\begin{equation}
S^i=S_0^i + \alpha\,.
\end{equation}
%%%%%
It is clear that if the original entropy product $\prod_i S_0^i$ were
quantised and
expressible purely in terms of the charges and angular momenta, then
the modified entropy product $\prod_i S^i$ would not be.
As we shall discuss in detail later, there is in fact a natural way
to remove the ambiguity in the definition of the entropy,  by requiring
that the black hole should have zero entropy in the case that its mass is sent
to zero.

    In section 2, we study the entropy product formulae for some
further examples
of charged rotating black holes in four and five dimensions that had not
previously been examined.  These include the four-dimensional
dyonic rotating black hole solutions of the Einstein-Maxwell-Dilaton
theory obtained by Kaluza-Klein reduction of pure five-dimensional gravity
\cite{Rasheed,LarsenKK},
and also the recently-constructed general three-charged rotating black holes of
five-dimensional gauged STU supergravity \cite{wu}.  We also consider
the charged
rotating black holes of four-dimensional $f(R)$-Maxwell theory
\cite{Larranaga:2011fv}.
Although $f(R)$ gravity is ostensibly a higher-derivative theory,
the known black-hole solutions have constant Ricci scalar $R$, and hence the
equations of motion are effectively reduced to second-order ones.

   Turning now to black holes in theories involving higher-derivative
gravity in a more non-trivial way, exact solutions
are rather harder to come by.
For higher-derivative theories whose Lagrangians are built from
polynomial curvature invariants, if each Riemann tensor is contracted
with at least two Ricci tensors then Einstein metrics continue to be solutions.
The Wald formula then implies that the entropy is proportional to the area of
the horizon, and the previous entropy-product formulae still hold for
any such black holes that are Einstein metrics.
Exact solutions
for static charged black holes have also
been found in Lovelock-Maxwell theory, and for these it was argued that
the entropy-product rule seemingly breaks down \cite{CDGK}.  However,
we argue
that this may just be an artefact of considering the rather degenerate
special case of non-rotating black holes.  The relevant point here,
as we shall discuss in detail later, is that
the total number of horizons for a black hole with generic non-vanishing
charges and angular momenta can be greater than the number of horizons in
special cases where charges and/or angular momenta vanish.  It is in the
generic case with the maximal number of horizons that one can expect the
product of horizon areas to be quantised.

  A simple illustrative example is provided by the Reissner-Nordstr\"om
solution,
which has two horizons located at the roots $r_\pm$ of $r^2-2Mr+Q^2=0$.
The product of the horizon areas is $A_+ A_-=(4\pi r_-^2)(4\pi r_+^2)=
 16\pi^2 Q^4$, which is indeed quantised and independent of the mass $M$.
Taking the limit when $Q$ goes to zero gives the area product $A_+ A_-=0$,
which, although trivial because of the factor $A_-=0$, is still quantised.
If, however, we were to consider the Schwarzschild solution in isolation, we
would say it has just one horizon, at $r_+=2M$, and the ``area product''
would simply be $A_+=4\pi r_+^2= 16 \pi M^2$, which is not quantised and
does depend on $M$. Thus one sees that the Schwarzschild black hole itself,
having only one rather than two horizons,
is not a sufficiently generic solution to reveal the underlying nature of
the quantised area-product formula for the Reissner-Nordstr\"om family.

  Returning to the black hole solutions of Lovelock-Maxwell theory
examined in \cite{CDGK}, it is quite plausible that the failure of the
area-product rule for the static black holes is again a
consequence of not considering the most generic situation, in this
case with rotation
included.  Unfortunately the more general rotating solutions in
the Lovelock-Maxwell
are not presently known, and so it is not possible at this time to settle
the question definitively.

    In section 3 we consider an example that is rather analogous, and
where we are able explicitly to illustrate
a similar phenomenon, namely for charged rotating black holes in
the conformally-invariant Weyl-Maxwell theory in four dimensions.
We demonstrate that in this example the entropy-product rule holds for
rotating black hole solutions but that it would fail if one
considered just the static non-rotating solutions in isolation.

In section 4, we comment on a general phenomenon for rotating black holes,
namely that if the metric is written as a timelike bundle over a
Euclidean-signature base space, with warp factors multiplying
the base and the fibre metrics, then the expression for the area of any
horizon is independent of the warp factor.  However, if one takes the
static limit, the area becomes dependent on the warp factor.
This observation could have further implications for the study of
the microscopic properties of general rotating black holes  in gravity
theories in diverse dimensions,  along the lines of  the
``subtracted geometry.''

We conclude our paper in section 5.

\section{Further Area-Product Examples in $D=4$ and $D=5$}

   In this section we consider three further examples of black hole
solutions in four and five dimensions for which area-product relations had
not previously been studied.  The first is the four-dimensional
rotating dyonic black hole solution of the Einstein-Dilaton-Maxwell theory
that can be obtained as the dimensional reduction of five-dimensional
pure gravity \cite{Rasheed,LarsenKK}.  Next, we look at the general
solution for a 3-charge
rotating black hole in five-dimensional gauged supergravity \cite{wu}.
The third example is the charged rotating black hole solution of Maxwell
theory coupled to $f(R)$ gravity in four dimensions \cite{Larranaga:2011fv}.

\subsection{Entropy product formula for the dyonic 
  KK black hole}\label{dyonicbh}

  The solution for the dyonic rotating Kaluza-Klein black hole carrying
electric and magnetic charge can be embedded in the four-dimensional
${\cal N}=2$ supergravity STU theory.  The electric and magnetic charges are
both carried by just one of the four gauge fields in the theory.
It should be noted that although
there exists a discrete duality symmetry in the KK reduction of 
five-dimensional gravity, 
under which the electric and
magnetic charge are exchanged, there is no continuous duality symmetry and
so this dyonic black hole cannot be rotated into a purely electric
or purely magnetic one.  Using
the notation and conventions of \cite{cclp}, the solution
can be written as
%%%%%
\be
d\hat s_4^2 = -e^{\varphi_4}\, (dt-\omega d\phi)^2 + e^{-\varphi_4}\, ds_3^2\,,
\label{d3red}
\ee
%%%%%
where
%%%%%
\bea
ds_3^2&=& (\rho^2-2mr)\Big(\fft{dr^2}{\Delta} +d\theta^2\Big) +
         \Delta\sin^2\theta\, d\phi^2\,,\nn\\
\Delta&=& r^2 + a^2-2mr\,,\qquad \rho^2=r^2+a^2\cos^2\theta\,.\label{kerr3}
\eea
%%%%%
The functions $\varphi_4$ and $\omega$ are given by
%%%%%
\bea
\omega &=& \fft{2am c_4\,[(r-m) \Xi + m c_5]\sin^2\theta}{\rho^2-2mr}\,,\nn\\
&&\nn\\
e^{2\varphi_4}&=& \fft{(\rho^2-2mr)^2\, \Xi}{(f_1 \Xi + 2m c_4^2 U_-)
   (f_2 + 2m \Xi U_+)}\,,
\eea
%%%%%
where we have defined
%%%%%
\bea
U_\pm &=& (r-m) c_5 \pm a s_4 s_5 \cos\theta\,,\quad
f_1 = \rho^2-2mr + 2m^2 c_4^2\,,\quad f_2=\rho^2-2mr + 2m^2 \Xi^2\,,\nn\\
c_i&=& \cosh\delta_i\,,\qquad s_i=\sinh\delta_i\,,\qquad
  \Xi=\sqrt{1+ c_4^2 s_5^2}\,.
\eea
%%%%%
The other non-vanishing fields in the four-dimensional supergravity theory
are
%%%%%
\bea
e^{2\varphi_1}&=& e^{2\varphi_2}=e^{2\varphi_3}=
    \fft{f_1 \Xi + 2m c_4^2 U_-}{\Xi\, (f_2 + 2m \Xi U_+)}\,,\nn\\
\hat A&=& \nu d\phi + \sigma_4 (dt-\omega d\phi)\,,\label{phi1A}
\eea
%%%%%
where
%%%%%
\bea
\nu &=& \fft{2m s_4 c_4 \Delta \cos\theta +
      2am c_4 s_5[c_4^2 c_5\, (r-m) + m \Xi] \sin^2\theta}{\Xi\, (\rho^2-2mr)}
  \,,\nn\\
\sigma_4 &=& \fft{2m^2 c_4^2 s_5 c_5 + 2m\Xi [(r-m) s_5 + a s_4 c_5\cos\theta]}{
   f_2 + 2m \Xi U_+}\,.
\eea
%%%%%

   There are two horizons, which are located at the two roots $r_\pm$
of $\Delta=0$.  The entropies are given by
%%%%%
\be
S_\pm = 2\pi m c_4\, \Big|r_\pm\, \Xi - m(\Xi-c_5)\Big|\,.\label{Spm}
\ee
%%%%%
(The absolute value must be used here because $r_-\, \Xi - m(\Xi-c_5)$
can be negative under appropriate conditions; see later.)
The angular momentum $J$ and the electric and magnetic charges $Q)$ and
$P$ are given by
%%%%%
\be
J = a m c_4 \Xi\,,\qquad Q=2m s_5 \Xi\,,\qquad P= \fft{2m s_4 c_4}{\Xi}\,,
\label{JQP}
\ee
%%%%%
from which it follows  that
%%%%%
\be
S_+ S_- = 4\pi^2 m^2 c_4^2\, \Big|a^2 \Xi^2 - m^2 s_4^2 s_5^2\Big|\,,
\ee
%%%%%
and hence, as can be seen from (\ref{JQP}),
%%%%%
\be
S_+ S_- = 4\pi^2\,\Big| J^2 - \fft1{16}\, P^2 Q^2\Big|\,.\label{dyonprod}
\ee
%%%%%

  It might, at first sight, seem surprising that with only one field
strength active, the charges contribute in the entropy product formula.
However, bearing in mind that the charge contribution in a general black hole
must be invariant under $SL(2,R)^3$, and that for the usual ``4-charge''
black hole the charge contribution is of the form $P_1 Q_2 P_3 Q_4$, this
is in fact correct.  See, for example, equation (6.22) in
\cite{duff} :\footnote{In \cite{duff}, the notation where a ``standard''
4-charge black hole has one electric charge $q_0$ and three magnetic
charges $(p^1, p^2, p^3)$ is used.  By contrast, in the notation of
\cite{cclp} the `standard'' 4-charge black hole has magnetic charges
$p^1$ and $p^3$, and electric charges $q_2$ and $q_4$.  Our presentation
of the Rasheed black hole has electric and magnetic charges $q_4$ and $p^4$.
In the notation of \cite{duff}, a simple choice for the Rasheed black hole
would be to take $q_0$ and $p^0$ non-zero, in which case the last term in
(\ref{duffform}) gives $-(q_0)^2 (p^0)^2$, in contrast to $+4 q_0 p^2 p^2 p^3$
for the standard 4-charge black hole \cite{CYII}.

See, also \cite{CT}, eq. (69),  where the manifestly ($S$- and $T$-) duality invariant quartic charge form was  first derived,  as it appeared in the entropy formula of  the most general BPS black hole   of four-dimensional ${\cal N}=4$ ungauged supergravity. For related U-duality invariant charge forms in four and five dimensions, see, e.g., \cite{CH}.
 }
%%%%%
\crampest
\bea
D(p,q)&=& 4[(p^1 q_1)(p^2 q_2) + (p^1 q_1)(p^3 q_3) +
  (p^2 q_2)(p^3 q_3) -p^0 q_1 q_2 q_3 +q_0 p^1 p^2 p^3] -(p^\mu q_\mu)^2\,.
\label{duffform}
\eea
\uncramp
%%%%%

For comparison, we may consider the standard 4-charge black hole.  In the
notation of \cite{cclp} this has magnetic charges $P_1$ and $P_3$, and
electric charges $Q_2$ and $Q_4$.  Evaluating the entropy-product
formula, we find
%%%%%
\be
S_+ S_- = 4\pi^2\Big( J^2 + \fft14 P_1 Q_2 P_3 Q_4\Big)\,.\label{4chprod}
\ee
%%%%%
This is indeed consistent with (\ref{dyonprod}) and, after
making the exchange $p^2\leftrightarrow q_2$, (\ref{duffform}).

\subsection{3-charge solution in $D=5$ gauged supergravity}

   Area-product formulae for a variety of rotating black holes in
gauged supergravities were studied in \cite{cvgipo}.  However, only more
recently has the general 3-charge rotating black hole in five-dimensional
gauged supergravity been constructed \cite{wu}, and so we are now in
a position to check the area-product relation for this example.
It has horizons determined by
the roots of a 6th-order polynomial in the radial coordinate $r$.  Since
only even powers of $r$ occur, we can define $x=r^2$ and reduce the problem
to one with three horizons, at the roots $x=x_1$, $x=x_2$ and $x=x_3$. The
radial function is then given by \cite{wu}
%%%%%
\bea
\Delta &=&(x+a^2)(x+b^2)(1+g^2 x) -2m x + 2m g^2\Big\{(s_1^2+s_2^2+s_3^2) x^2
\nn\\
&& -(s_1^2 s_2^2 + s_1^2 s_3^2 + s_2^2 s_3^2)[(a^2+b^2-2m) x +
  a^2 b^2 (2+g^2 x)]\nn\\
&& + s_1^2 s_2^2 s_3^2\Big([(a+b)^2 -2m)][(a-b)^2-2m] -
  2g^2 a^2 b^2 (2x+2m+a^2+b^2) + g^4a^4 b^4\Big)\nn\\
&&+ 2 mg^2 a^2 b^2 [s_1^4 s_2^4 + s_1^4 s_3^4 + s_2^4 s_3^4 -
   2 s_1^2 s_2^2 s_3^2(s_1^2+s_2^2+s_3^2)]\Big\}\,.
\eea
%%%%%
The entropy, angular momenta and charges are also given in
\cite{wu}.  After transforming to the variable $x=r^2$, the entropy at
the $i$'th horizon is given by
%%%%
\be
S_i =\fft{\pi^2}{2\chi_a\, \chi_b}\sqrt{\fft{W_i}{x_i}}\,,\label{SW}
\ee
%%%%%
where
%%%%%
\bea
W_i &=& \big[(x_i+a^2)(x_i+b^2) +2mx_i(s_1^2+s_2^2+s_3^2)\big]
\Big\{(x_i+a^2)(x_i+b^2) \nn \\
&& +2mg^2[(a+b)^2 -g^2a^2b^2][(a-b)^2 -g^2a^2b^2]s_1^2s_2^2s_3^2 -4mg^2a^2b^2(s_1^2s_2^2 \nn \\
&& +s_1^2s_3^2+s_2^2s_3^2)\Big\} +8m^2 x_i c_1c_2c_3s_1s_2s_3ab\sqrt{\Xi_{1a}\Xi_{2a}\Xi_{3a}\Xi_{1b}\Xi_{2b}\Xi_{3b}} \nn \\
&& +4m^2x_i\big[x_i+g^2a^2b^2(s_1^2+s_2^2+s_3^2)\big](s_1^2s_2^2+s_1^2s_3^2+s_2^2s_3^2) \nn \\
&& -4m^2\Big\{(a^2+b^2)(1+g^2a^2)(1+g^2b^2)x_i +g^2\big[(a^4+b^4)x_i \nn \\
&& +g^2a^4b^4(2+g^2 x_i)\big](s_1^2+s_2^2+s_3^2) +g^2a^2b^2(a^2+b^2 \nn \\
&& +g^2a^2b^2)\big[2+g^2x_i(s_1^2s_2^2+s_1^2s_3^2 +s_2^2s_3^2)\big]\Big\}s_1^2s_2^2s_3^2 \nn \\
&& +4m^2g^4a^4b^4(s_1^4s_2^4+s_1^4s_3^4+s_2^4s_3^4) -8m^2 x_i g^6a^4b^4s_1^4s_2^4s_3^4 \nn \\
&& +8m^3(x_i+g^2a^2b^2)s_1^2s_2^2s_3^2 \,,
\eea
%%%%%
and
%%%%%
\be
\chi_a=1-g^2 a^2\,,\quad \chi_b=1-g^2 b^2\,,\quad
\Xi_{ia}= 1 + g^2 a^2 s_i^2\,,\quad \Xi_{ib}=1+g^2 b^2 s_i^2\,.
\ee
%%%%%
The angular momenta and electric charges are given by \cite{wu}
%%%%%
\bea
J_a &=&
\frac{\pi m}{2\chi_a^2\chi_b}\big(ac_1c_2c_3\sqrt{\Xi_{1a}\Xi_{2a}\Xi_{3a}}
 -b\chi_a^2s_1s_2s_3\sqrt{\Xi_{1b}\Xi_{2b}\Xi_{3b}}\big) \, , \\
J_b &=& \frac{\pi m}{2\chi_a\chi_b^2}\big(bc_1c_2c_3\sqrt{\Xi_{1b}\Xi_{2b}\Xi_{3b}}
 -a\chi_b^2s_1s_2s_3\sqrt{\Xi_{1a}\Xi_{2a}\Xi_{3a}}\big)\,,\nn\\
Q_i &=& \frac{\pi m}{2\chi_a\chi_b}
\Big(\frac{c_is_i\sqrt{\Xi_{1a}\Xi_{2a}\Xi_{3a}\Xi_{1b}\Xi_{2b}\Xi_{3b}}}{\sqrt{\Xi_{ia}\Xi_{ib}}}
 -g^2ab\frac{c_1c_2c_3s_1s_2s_3}{c_is_i}\sqrt{\Xi_{ia}\Xi_{ib}}\Big) \,.
\eea
%%%%%%

  From these it is a straightforward matter to compute the
product of the entropies, and express it in terms of the conserved
charges.  We find that it can be written as
%%%%%
\be
\prod_{i=1}^3 S_i =
  \pm\fft{2\im \pi^3}{g^3}\,( \pi J_a J_b + 4 Q_1 Q_2 Q_3)\,.
\ee
%%%%%
This result reduces to the special cases presented previously in
\cite{cvgipo} if two or more charges are set equal.
(The $\pm$ sign on the right-hand side reflects the fact that there is
a sign ambiguity in taking the square roots in (\ref{SW}).
This would not be seen if
we took the product over all six horizons in the language of the
original radial variable $r$, in which case the right-hand side would be
squared.)

\subsection{Charged rotating black hole in $f(R)$ theory}

Although $f(R)$ gravities involve in general higher derivatives in their
equations of motion, their solutions include those for which
the Ricci scalar is constant.  In this case, the effective equations of
motion are then reduced to second order.  In four dimensions, it happens
that the trace of the Maxwell energy-momentum tensor
vanishes, and so one can still construct analytic charged solutions in this case
for which $R$ is constant.

  The four-dimensional Lagrangian for $f(R)$ gravity coupled to a
Maxwell field is given by
%%%%%
\begin{equation}
{\cal L}_4=\sqrt{-g} (f(R) - \ft14 F^2)\,.
\end{equation}
%%%%%
The Einstein equations of motion are then
%%%%%
\begin{equation}
f'(R)R_{\mu\nu} -\ft12 g_{\mu\nu} f(R) +
(\nabla_\mu\nabla_\nu - g_{\mu\nu}\Box) f'(R) =
\ft12 (F^2_{\mu\nu} - \ft14 g_{\mu\nu} F^2)\,,\label{einsteineq}
\end{equation}
%%%%%
where $f'(R)$ means $\del f(R)/\del R$.
The trace of the Einstein equation is independent of the Maxwell field,
%%%%%
\begin{equation}
f'(R)R - 2 f(R)-3\Box f'(R) =0\,,
\end{equation}
%%%%%
and so it admits solutions where the Ricci scalar is constant, $R=R_0$,
where
%%%%%
\begin{equation}
f'(R_0)R_0=2f(R_0)\,.
\end{equation}
%%%%%
For solutions with $R=R_0$, the Einstein equations (\ref{einsteineq})
then reduce to
%%%%%
\begin{equation}
R_{\mu\nu} - \ft12 g_{\mu\nu} (R - 2\Lambda) =
\ft12 (\widetilde F_{\mu\nu}^2 - \ft14 g_{\mu\nu} \widetilde F^2)
\,,\label{kneom}
\end{equation}
%%%%%
where $\Lambda=\fft14 R_0$ is the effective cosmological constant and
$\widetilde F_{\mu\nu} = F_{\mu\nu}/\sqrt{f'(R_0)}$.  This is precisely
the same equation as in Einstein-Maxwell theory, which admits
the well-known Reissner-Nordstr\"om and Kerr-Newman solutions,
except that now the charge is scaled by the $1/\sqrt{f'(R_0)}$ factor.
This gives rise to the static \cite{Moon:2011hq} and
rotating \cite{Larranaga:2011fv} charged black holes in four-dimensional $f(R)$ gravity coupled to the Maxwell field.  Note that the entropy is
also given by one quarter the area of the horizon, scaled by the
$f'(R_0)$ factor.  Thus the entropy-product formula for the charged
rotating black hole in $f(R)$ gravity is similar to that for the
Kerr-Newman solution in Einstein-Maxwell theory, but for an overall scaling
by an $f'(R_0)$-dependent factor.  The entropy product formula for the static case was discussed in \cite{CDGK}.

\section{Entropy Product Formula for Maxwell-Weyl Theory}

In four dimensions, owing to the fact that Gauss-Bonnet term is a
total derivative, the most general quadratic-curvature Lagrangian
can be parameterised as
%%%%%
\begin{equation}
{\cal L}_2=\alpha R^\mu\nu R_{\mu\nu} + \beta R^2\,.\label{quad}
\end{equation}
%%%%%
Any Einstein metric with cosmological constant $\Lambda$, which is
a solution for the theory described by the Lagrangian
${\cal L}_0 = R -2 \Lambda$, will continue to be a solution of
the theory described by
${\cal L}={\cal L}_0 + {\cal L}_2$.  For any such black hole solution,
the Wald formula implies that the entropy will just be a
constant multiple of the area of the horizon, and hence entropy-product
results for the Kerr-AdS metric in Einstein gravity will continue
to holds in the extended theory. If black hole solutions over and above
those that are Einstein metrics existed, then their entropy-products would
need to be investigated in their own right.

  In fact here no explicit black hole solutions, beyond Kerr-AdS, are
known for the general case of cosmological Einstein gravity augmented by the
quadratic-curvature Lagrangian (\ref{quad}).  The situation becomes simpler,
however, if we consider pure conformal gravity, where the Lagrangian is
simply a multiple of the square of the Weyl tensor, and additional
non-Einstein black hole solutions can be found.  In fact non-trivial
such solutions can also be found in the conformally-invariant Weyl-Maxwell
theory, described by the Lagrangian
%%%%%
\begin{equation}
{\cal L} = \sqrt{-g}\Big(\ft12 \alpha C^{\mu\nu\rho\sigma}
   C_{\mu\nu\rho\sigma} +
\ft13\alpha  F^2\Big) = \sqrt{-g}\Big(\alpha R^{\mu\nu} R_{\mu\nu} -
\ft13\alpha R^2 +
\ft13\alpha  F^2\Big) + \alpha {\cal L}_{\rm GB}\,.
\end{equation}
%%%%%
(Here ${\cal L}_{\rm GB}$ denotes the Gauss-Bonnet Lagrangian.)

\subsection{Charged rotating black holes}

Charged rotating black holes in the four-dimensional conformally-invariant
Einstein-Weyl theory
%%%%%
\be
{\cal L} =\sqrt{-g}\Big( \fft12 \alpha C^{\mu\nu\rho\sigma} C_{\mu\nu\rho\sigma}
+\fft13 \alpha F^{\mu\nu} F_{\mu\nu}\Big)
\ee
%%%%%
were studied in \cite{liulu}.  The solution for a dyonic black hole can be
written as \cite{liulu}
%%%%%
\bea
ds_4^2 &=& \rho^2\Big(\fft{dr^2}{\Delta_r} +
  \fft{d\theta^2}{\Delta_\theta}\Big) +
 \fft{\Delta_\theta \sin^2\theta}{\rho^2}\Big(a dt -
  (r^2+a^2)\fft{d\phi}{\Xi}\Big)^2 -
  \fft{\Delta_r}{\rho^2}\Big(dt - a \sin^2\theta \fft{d\phi}{\Xi}\Big)^2
 \,,\nn\\
A &=& \fft{q r}{\rho^2}\, \Big(dt - a \sin^2\theta \fft{d\phi}{\Xi}\Big) +
  \fft{p\cos\theta}{\rho^2} \Big(a dt -
  (r^2+a^2)\fft{d\phi}{\Xi}\Big)\,,
\eea
%%%%%
where
%%%%%
\bea
\rho^2&=& r^2 + a^2\cos^2\theta\,,\qquad \Delta_\theta= 1 -g^2 a^2 \cos^2\theta
\,,\qquad \Xi= 1-g^2 a^2\,,\nn\\
\Delta_r &=& (r^2+a^2)(1+g^2 r^2) -2m r +\fft{(p^2+q^2) r^3}{6m}\,.
\eea
%%%%%
The horizons occur at the roots of $\Delta_r=0$.
In what follows we shall set $p=0$ so that there is only an electric
charge, since the inclusion of a magnetic charge adds no further features
of relevance to the discussion.

  The conserved energy, charge and angular momentum are given by \cite{liulu}
%%%%%
\be
E= \fft{2\alpha g^2}{\Xi^2}\Big(m + \fft{a^2 q^2}{12 m}\Big)\,,\qquad
Q = \fft{\alpha q}{3\Xi}\,,\qquad
J= \fft{2a \alpha g^2}{\Xi^2}\Big(m +\fft{q^2}{12 m g^2}\Big)\,,
\ee
%%%%%
and the Wald entropy at a root $r_i$ is given by
%%%%%
\be
S_i= \fft{2\pi\alpha}{\Xi}\Big( 1 + g^2 r_i^2 +\fft{q^2 r_i}{6m} - c\,
 \Xi\Big)\,,
\ee
%%%%%
The constant $c$ is purely numerical (i.e. parameter
independent), and corresponds to adding a constant multiple of the
Gauss-Bonnet invariant to the action.  If we choose $c=1$, the
resulting Lagrangian involves only the Ricci tensor and Ricci scalar,
and we have
%%%%%
\begin{equation}
S_i = \fft{2\pi \alpha}{\Xi} \Big( \fft{g^2}{4\pi} A_i +
          \fft{q^2 r_i}{6m}\Big)\,,
\end{equation}
%%%%%
where $A_i=4\pi (r^2_i + a^2)$ is the area of the $i$'th ``horizon.''  This
is a more natural definition for the entropy of the system since the
entropy vanishes when the solution becomes the vacuum. Calculating the
product of the entropies at the four horizons, keeping the constant $c$
arbitrary for now, and
then expressing the result in terms of the conserved charges,
we find
%%%%%
\bea
\prod_{i=1}^4 S_i &=& (2\pi)^4\alpha^2\Big[c J^2 + (1-c) E^2 g^{-2} + 3c(1-c) Q^2
+ c^2(1-c)^2 \alpha^2\Big]\,.
\eea
%%%%%
Making the natural choice $c=1$ discussed above yields the result
%%%%%
\be
\prod_{i=1}^4 S_i = (2\pi)^4\alpha^2 J^2\,,
\ee
%%%%%
which is indeed quantised.  This is a highly non-trivial result since now the entropy is no longer simply one quarter of the area of the horizon, in which case, the quantisation of the entropy product implies the geometric quantisation of the product of all horizon areas.

\subsection{Charged static black holes}

In the previous subsection, we saw that with the natural choice $c=1$
the product of entropies depended
only on the angular momentum, but was independent of the charges.
If we send the angular momentum to zero, the entropy of one of the
horizons becomes also goes to zero.  More precisely, in the static case,
the four null surfaces are reduced to only three.  There are in
fact more general static solutions than the obvious one resulting
from setting $a=0$. The most general static solution is given by
\cite{Riegert:1984zz}
%%%%%
\begin{eqnarray}
ds^2 &=& -f dt^2 + \fft{dr^2}{f} + r^2 d\Omega_2^2\,,\qquad
A=-\fft{q}{r} dt\,,\cr
f&=& -\ft13 \Lambda r^2 + c_1 r + c_0 + \fft{d}{r}\,,
\qquad 3c_1 d + 1 + q^2=c_0^2\,,
\end{eqnarray}
%%%%%
and the entropies are given by
%%%%%
\begin{equation}
S_i = -\fft{2\pi \alpha  (3d + (c_0+2) r_i)}{3r_i}\,.
\end{equation}
%%%%%
Thus we have
%%%%%
\begin{equation}
\prod_{i=1}^3 S_i = -\ft{8}{27} \pi^3 \alpha^3 (2 - 3c_0 + c_0^3 -
26 d^2 g^2) + \ft{128}{9} \alpha (c_0+2) \pi^3 Q_e^2\,,
\end{equation}
%%%%%
where $Q_e=\fft14\alpha q$ is the electric charge.  It is thus clear
that the product of entropies is longer expressed purely in terms of
quantised charges in this static case.

      An important lesson one learns from the Weyl-Maxwell theory is that
in higher-derivative gravities, the failure of the entropy product formula
in the static case does not necessarily imply the failure of the formula
in the more general rotating solutions with angular momenta as well.

      After all, for
Schwarzschild black holes with or without a
cosmological constant, the entropy or the product of entropies depends on
the mass, rather than on any quantised quantities.  However, as we
discussed for the Schwarzschild black hole in the introduction,
the static
solution is a special case of the more general solutions with angular
momenta or charges, which have a larger number of null surfaces.  The
static solution corresponds to degenerate limit where the area of
one or more of the null surfaces of the more general class of solutions
goes to zero.

     In all the examples examined so far, as long as maximally rotating
solutions exist in two-derivative or higher-derivative gravities,
the entropy product formulae do work, in the sense of depending only on the
products of angular momenta with (or without) charges.  However in many cases, such
as in Lovelock gravities, the exact solutions for rotating black holes are
unknown. We expect that the entropy-product formulae will work in these
cases, even if they fail for the known, but rather degenerate,
charged static solutions.

\section{Warp-factor Independence and the Static Limit}

   Most of the charged black hole metrics in ungauged supergravities were
constructed by using solution-generating techniques, starting from a
an uncharged black hole as a ``seed'' solution.  One of the most
universal solution-generating techniques involves performing a
dimensional reduction of the seed metric to three dimensions, and then
acting with global symmetries of the associated
non-linear sigma model coupled to gravity in three dimensions.  For example,
in the case of constructing charged rotating solutions in four dimensions, one
performs a timelike reduction using the Kaluza-Klein ansatz given in
(\ref{d3red}). The reduced three-dimensional metric $ds_3^2$ is invariant
under the global symmetry transformations, and thus remains the same
as in the original reduction of the seed Kerr solution, as in (\ref{kerr3}).
The specific forms of the warp factor $e^{\varphi_4}$ and the function
$\omega$ in (\ref{d3red}) will depend on the details of the theory under
consideration, and the nature of the charges that are turned on, but the
general structure, and the universality of the 3-metric $ds_3^2$, will be
common to all examples.

  The horizons of the charged metric will be located at the same radii
$r_i$ as those of the original seed metric, namely at the zeroes of
the function $\Delta$.  It is then evident from (\ref{d3red}) and
(\ref{kerr3}) that the area of the horizon at $r=r_i$ will be given by
%%%%%
\be
{\cal A}_i = 4\pi \sqrt{(2mr -\rho^2)\omega^2}\Big|_{r=r_i}\,,\label{stationary}
\ee
%%%%%
and so, in particular, it is independent of the warp factor $e^{\varphi_4}$.

   The above discussion assumes that the metric is stationary, but not static.
In the static case, $a=0$ and the function $\omega$ vanishes, and so
evidently for the
metric to be non-degenerate at the horizons it must be that the warp
factor acquires a factor of $\Delta$ that can cancel the overall
factor of $\Delta$ in the 3-metric $ds_3^2$.  The area of the horizon at
$r=r_i$ is now given by
%%%%%
\be
{\cal A}_i = 4\pi \Big[ e^{-\varphi_4}\, \Delta\Big]_{r=r_i}\,,
\ee
%%%%%
which, unlike in the rotating case, {\it does} depend upon the warp factor
$e^{\varphi_4}$.

The independence of the horizon area on the warp factor for stationary
black-hole metrics has been noted
in earlier works, and indeed it has formed the basis for the notion of
``subtracted geometries'' that was considered in \cite{CL11I,CL11II}.  In
those papers, the proposal was to subtract certain terms in the
warp factors of  the black hole metrics, in such a way that the massless scalar wave equation for the subtracted geometry  attains a manifest $SL(2,{\bf R})\times SL(2,{\bf R})$ conformal symmetry.
It was noted in \cite{CL11I,CL11II} that  the areas of the two horizons and 
the periodicity of the azimuthal angles for the rotating solutions are 
unchanged in the subtracted geometries.  It could, however, be troubling 
if the phenomenon we have noted above, in which the warp factor does enter 
in the area formula in the static case, were to signal a discontinuity if 
one approached the static situation as an $a\rightarrow 0$ limit of the 
rotating solution.  It is of interest,
therefore, to investigate this limit in detail in explicit examples.

   We saw in section \ref{dyonicbh} that the entropy-product formula
(\ref{dyonprod}) for the rotating dyonic black hole required an
absolute value of the $J^2-P^2 Q^2/16$ factor on the right-hand side, to
allow for the case where $J^2<P^2 Q^2/16$.  Specifically, the origin of
this non-analytic dependence on the conserved charges is that the metric
function $\omega$ in (\ref{stationary}) changes sign and becomes
negative at the inner horizon if $J^2$ becomes less than $P^2 Q^2/16$.  A
direct calculation of the entropy product for the static metric with $a=0$
confirms that indeed
%%%%%
\be
S_+ S_- = \fft14 \pi^2 P^2 Q^2\,.\label{static}
\ee
%%%%%
Thus we see in this example that the direct evaluation in the static metric
is in agreement with
the result obtained by taking a $J\rightarrow 0$ limit.

   It is noteworthy that it is possible to cast the metric in the
warped-product form (\ref{d3red}) for
any axisymmetric solution in four dimensions. In five dimensions, all 
known rotating black-hole
solutions in gauged and ungauged supergravity can be cast in the
analogous form:
%%%%%
\be
ds_5^2 = -\Delta^{-1/3}(dt+\omega_\phi d\phi + \omega_\psi d\psi)^2 +
   \Delta^{2/3}ds_4^2\,,
\ee
%%%%%
where  the four-dimensional base metric $ds_4^2$ is  K\" ahler, and the  
warp  factor $\Delta$ depends only on the radial coordinate $r$  and the 
polar angle $\theta$. This is the case for all known rotating black holes, 
including \cite{BC} the  general rotating  three charge rotating AdS 
black hole that was obtained in \cite{wu}.
We have checked  for many of these examples  that  their  entropy is 
independent of the warp-factor.

\section{Conclusions}

     In this paper, we discussed the universal nature of the
product of the entropies associated with all the horizons of a black hole,
and in particular that the product depends only on the charges and angular
momenta, which are subject to quantisation conditions.  The entropy-product
formulae are valid as long as either the maximum number of rotation
parameters, and/or the maximum number of charges, are turned on.
The meaning of the former is clear, namely that there is a non-vanishing
rotation in each of the $[(D-1)/2]$ orthogonal spatial 2-planes. 
The notion of the ``maximum number of
charges'' requires further explanation.  It is well known that
Einstein-Maxwell gravity in four or five dimensions, with or without a
cosmological constant, can be embedded in a four or five-dimensional
supergravity.  The Reissner-Nordstr\"om black hole solution in the
associated supergravity
theory can be viewed as a superposition of some more basic $U(1)$-charged
building blocks.  For example, four-dimensional Einstein-Maxwell gravity
can be embedded in the STU model, which is a consistent 
truncation of maximally supersymmetric ungauged supergravity that has
${\cal N}=2$ supersymmetry and four $U(1)$ gauge fields. The quantised 
entropy-product formula is valid provided
that all
four charges are turned on, with the Reissner-Nordstr\"om  black hole then
corresponding to
the special case where the four charges are set equal.  

   This picture can be
extended to higher-dimensional non-supersymmetric theories, where the
Reissner-Nordstr\"om black hole again emerges as a superposition of more
fundamental ingredients \cite{HL}.  Thus for a theory that supports
multiply-charged black holes, if the solution specialises to 
the Reissner-Nordstr\"om black hole when the charges are equated, then the 
number of charges can be
viewed as ``maximal,'' and the entropy-product formulae will hold.
In the dyonic black hole we discussed in section 2, although it involves
only two charges, namely the electric and magnetic charges carried by a
single $U(1)$ gauge field in the STU model, rather than
the usual four charges carried by all four gauge fields, it still can
be viewed as having the maximum number of charges since the solution can
be reduced to the Reissner-Nordstr\"om black hole if the charges are 
equated.\footnote{Note that in ungauged supergravity theories in four and  
five dimensions, the product of horizon areas is governed, respectively, 
by quartic or cubic charge-forms, which are modulus independent and 
invariant under duality transformations. (See,  for example,  \cite{CT}  
for the quartic charge form  in ${\cal N}=4$  ungauged supergravity).  
Note that in four dimensions both the four-charge black holes  \cite{CYII} 
and the Kaluza-Klein black hole \cite{Rasheed,LarsenKK} are special 
examples where the duality invariant quartic charge form is non-zero.} (As 
discussed in the appendix, for dilatonic AdS black holes in dimensions 
$D\ge 6$, the entropy-product rule can work in cases where there are fewer 
than the maximal number of charges.)

   The situation changes in higher-derivative gravity.   The
changes are two-fold:  Firstly, in general higher-derivative gravity, 
the entropy 
may no longer be proportional to the area of the horizon, and thus the 
quantisation of the product of entropies may no longer be a purely geometric 
property.  Secondly, the concept of ``maximal charges'' has to be refined.
In the explicit example of the conformally-invariant Maxwell-Weyl gravity, 
we demonstrated that the entropy-product formula still works, but the result 
is expressed in terms of  the angular momentum only,  with no 
dependence on the charge.  This implies that the Maxwell charge is no 
longer ``maximal,''
in the sense discussed above.  In order to obtain an entropy-product
formula that is quantised, one must therefore necessarily 
turn on the angular momentum.

  The example of charged static black holes in another higher-derivative
theory, Maxwell-Lovelock gravity, was studied in \cite{CDGK}, and it was
found there that the product of the entropies did not satisfy a
quantisation rule.  This suggests that the phenomenon we found in
Maxwell-Weyl gravity may be more widespread in higher-derivative
theories; one must have non-zero angular momentum in order to
obtain a quantised entropy-product formula.

  Of course, the higher-derivative examples that we have been discussing all
have in common that the gravitational action is higher-derivative,
while the Maxwell gauge field action is left unaltered.
A natural conjecture, therefore, is that in higher-derivative theories
the Maxwell field should have appropriate higher-derivative terms also, in
order to acquire the ``maximal'' status.  In fact, such terms are
natural in higher-derivative extensions of supergravities.  The success
of the entropy product formulae for the rotating black hole in
Maxwell-Weyl gravity suggests this possibility.  In a spherical
dimensional reduction,
the angular momentum can be viewed as the electric charge of some
Kaluza-Klein vector associated with an isometry of the sphere.
The reduction of such a higher-derivative theory will clearly
give rise to a Kaluza-Klein vector with higher-derivative terms also.

To conclude, there appears to be a robust rule that the product of
entropies at all horizons of a black hole can be expressed purely in terms
of quantized quantities provided that the maximum numbers of angular
momenta and/or charges are turned on.  The key point in all cases is that
the maximal number of distinct ``horizons'' should be attained.
In two-derivative theories, the
notion of ``maximum number of charges''  can be restated as the condition
that by equating them appropriately, the solution can be reduced to
the Reissner-Nordstr\"om black hole.  In higher-derivative theories,
the notion is  not yet entirely clear, and is worthy of further investigation.

\section*{Acknowledgements}

M.C.~and H.L.~are grateful to the George and Cynthia Woods Mitchell
Institute for Fundamental Physics and Astronomy at Texas A\&M University
for hospitality during the early stages of this work.  M.C.
thanks the Department of Physics at Beijing Normal University, were initial ideas were discussed,  for hospitality and the Theory Division of CERN for hospitality during completion of
the project. H.L.~is grateful
to ICTS-USTC for hospitality during the later stages of this work. We are
grateful to Gary Gibbons for useful discussions.
M.C. would like to thank  Tolga Birkandan, Monica Guica, Finn Larsen and Zain Saleem for extensive discussions and collaborations on related topics.
M.C. is supported in part by DOE grant DE-SC0007901, the Fay R. and
Eugene L. Langberg Endowed Chair and the Slovenian Research Agency (ARRS).
The research of H.L.~is supported in part by the NSFC grants 11175269
and 11235003.  C.N.P.~is supported in part by DOE grant DE-FG02-95ER40917.

\appendix

\section{On Maximal Charges}

  Earlier in the paper we introduced the concept of 
a ``maximal'' number of charges in a charged black-hole solution.  
The concept is mostly simply discussed in the context of static solutions.  
In general, charged black holes are constructed as superpositions of 
multiple charges associated with different vector gauge fields in the theory. 
If the solution reduces to 
the Reissner-Nordstr\"om black hole under an appropriate specialisation of
the charges, then we refer to such multiply-charged black holes as 
having the ``maximum number'' of charges.  For these solutions, the 
entropy-product at all horizons is expressed in terms of charges only. 
However, the converse is not necessarily true.  For example, gauged 
supergravities in seven or six dimensions cannot be truncated 
to Einstein-Maxwell theory, and hence the charged AdS black holes cannot 
be reduced to Reissner-Nordstr\"om-AdS black holes by specialising the
charges.  Nevertheless, for charged black holes in these theories, 
built from two basic 
ingredients where one or the other of two independent gauge fields is
excited, the entropy-product rule still holds.   In this appendix, we 
shall comment further on these observations.

    Let us consider the Lagrangian proposed in \cite{HL} as the focus for
our discussion, since the theory contains all the essential properties of 
supergravities in terms of the structure of its relevant solutions. The 
Lagrangian is given by
%%%%%
\begin{eqnarray}
e^{-1} {\cal L}_D &=& R - \ft12 (\partial\phi)^2 - 
\ft14 e^{a_1\phi} F_1^2 - \ft14 e^{a_2\phi} F_2^2 - V(\phi)\,,\label{lag1}\\
V(\phi) &=& -\fft{g^2N_1}{4} \Big[2(D-3)^2 (N_1-1) e^{-a_1\phi} + 
2 a_1^2 (D-3)(D-2) N_1 e^{-\fft12(a_1+a_2)\phi}\cr
&&\qquad - a_1^2(D-2)\Big((D-3)N_1 - (D-1)\Big) 
e^{-a_2\phi}\Big]\,,\label{scalarpot}
\end{eqnarray}
%%%%%
where the constants $(a_1, a_2)$ satisfy the constraint
%%%%%
\begin{equation}
a_1 a_2 = - \fft{2 (D-3)}{D-2}\,.\label{a1a2cons}
\end{equation}
%%%%%
The theory admits charged AdS black holes \cite{HL}, given by
%%%%%
\begin{eqnarray}
ds^2 &=& -(H_1^{N_1} H_2^{N_2}) ^{-\fft{(D-3)}{D-2}} f dt^2 + 
(H_1^{N_1} H_2^{N_2})^{\fft{1}{D-2}} (f^{-1} dr^2 + r^2 d\Omega_{D-2}^2)\,,\cr
A_1&=& \ft{\sqrt{N_1}\,c_1}{s_1}\, H_1^{-1} dt\,,\qquad
A_2 = \ft{\sqrt{N_2}\, c_2}{s_2}\, H_2^{-1} dt\,,\cr
\phi &=& \ft12 N_1 a_1 \log H_1 + \ft12 N_2 a_2 \log  H_2\,,\qquad
f=1 - \fft{\mu}{r^{D-3}} + g^2 r^2 H_1^{N_1} H_2^{N_2}\,,\cr
H_1&=&1 + \fft{\mu s_1^2}{r^{D-3}}\,,\qquad
H_2 = 1 + \fft{\mu s_2^2}{r^{D-3}}\,,\label{solution3}
\end{eqnarray}
%%%%%
where $s_i=\sinh\delta_i$, $c_i=\cosh\delta_i$, and $(N_1, N_2)$ are 
given by
%%%%%
\begin{equation}
N_1 + N_2 = \fft{2(D-2)}{D-3}\,,\qquad
a_1^2 = \fft{4}{N_1} - \fft{2(D-3)}{D-2}\,.
\end{equation}
%%%%
The solution becomes the Reissner-Nordstr\"om-AdS black hole 
if we set $\delta_1=\delta_2$.

The thermodynamic quantities are all calculated in \cite{HL}.  For our 
purposes, we shall give only the charges and entropy:
%%%%%
\begin{equation}
Q_i=\fft{(D-3)\omega_{\sst{D-2}}}{16\pi} \mu N_i c_i s_i\,,\qquad
S=\ft14 r_0^{D-2} H_1(r_0)^{N_1/2} H_2(r_0)^{N_2/2} \Omega_{D-2}\,.
\end{equation}
%%%%%
For the case when $g=0$, the general solution has two real horizons;
the outer horizon is located at $r_0=\mu^{1/(D-3)}$, and the inner horizon is 
at $r=0$.  The entropy-product formula, ignoring inessential numerical 
constants, is given by \cite{HL}
%%%%%
\begin{equation}
S_+ S_-\sim Q_1^{N_1} Q_2^{N_2}\,.
\end{equation}
%%%%%
If we turn off one of the charges, then $r=0$ is no longer a horizon, but 
a singularity with zero area.  Thus this example demonstrates the validity 
of our definition of ``maximal'' charges.

   For non-vanishing $g$, the situation can be more subtle.   The 
horizons are located at all the roots of the metric function $f$.  
For rational values of $N_i$, the number of roots is finite.  The general 
formula determining these roots is rather involved.  For the 
Reissner-Nordstr\"om-AdS black holes, it can be shown that
%%%%%
\begin{equation}
\prod_i S_{i=1}^{2(D-2)}\sim (g^{-1}Q)^{2(D-2)}\,.
\end{equation}
%%%%%
It was conjectured in \cite{HL} that the general product formula, for
unequal charges, is
%%%%%
\begin{equation}
\prod_{i=1}^{2(D-2)} S_i\sim [(g^{-1}Q_1)^{N_1}(g^{-1} Q_2)^{N_2}]^{D-3}\,.
\end{equation}
%%%%%
The entropy-product formula appears to be suggesting that it fails to work
if we set one charge to zero, say $Q_2=0$.  This is clearly the case for $g=0$, as we have discussed.  However, if we let $N_1=2$ and $Q_2=0$, we find that for $D\ge 6$, the entropy formula continues to work:
%%%%%
\begin{equation}
\prod_{i=1}^{2(D-3)} S_i \sim (g^{-1} Q_1)^{2(D-3)}\,,\qquad \hbox{for} \qquad D\ge 6\,,\label{lessmaximum}
\end{equation}
%%%%%
that is to say, the product of entropies depends only on the charges and the
gauge coupling constant $g$.
Note that the total number of horizons is reduced from $2(D-2)$ to $2(D-3)$. On the other hand, the entropy-product formula indeed fails to work when $D=4$ 
or 5 with this charge specification.  In the cases of $D=4$, 5, 6 or 7, the 
theory can be embedded in a gauged supergravity and the results were 
found already in \cite{cvgipo}.  

   It is somewhat surprising that when we
 have less than the maximal number of charges, the entropy-product formula 
still works.  The most likely explanation can be found in the superposition 
rule
%%%%%
\begin{equation}
N_1 + N_2 = \ft{2(D-2)}{D-3}\,.
\end{equation}
%%%%%
This shows that $N_i$ can only take integer values in $D=4$ and 5.  In 
dimensions higher than five, the largest integer value that $N_1$ can take 
is $N_1=2$, in which case $N_2$ is fractional and  smaller  than 1.  
In other words, $N_1=2$ is the maximum integer ingredient, and that seems 
to be sufficient for the entropy-product rule to work, even if $Q_2$ 
is turned off.  Of course, it is worth emphasising  again that for the
 ``ungauged''  supergravity theories with $g=0$, the entropy-product 
formula requires that both charges $Q_i$ are turned on.


\begin{thebibliography}{99}

\bibitem{SV}
A. Strominger and C. Vafa,
{\it Microscopic origin of the Bekenstein-Hawking entropy},
Phys. Lett. {\bf B379}, 99 (1996), hep-th/9601029.
%%CITATION = PHLTA,B379,99;%%

\bibitem{Sen}
  A.~Sen,
{\it Black hole entropy function, attractors and precision counting of
  microstates,}
  Gen.\ Rel.\ Grav.\  {\bf 40}, 2249 (2008),
  arXiv:0708.1270 [hep-th].
  %%CITATION = GRGVA,40,2249;%%

\bibitem{CYII} M. Cveti\v c and D. Youm,
{\it Entropy of non-extreme charged rotating black holes in string theory},
Phys. Rev. {\bf D54}, 2612 (1996), hep-th/9603147.
%%CITATION = HEP-TH/9603147;%%

\bibitem{CYI}
 M. Cveti\v c and D.~Youm,
{\it General rotating five dimensional black holes of toroidally compactified
heterotic string},
Nucl. Phys. {\bf B476}, 118 (1996), hep-th/9603100.
%%CITATION = NUPHA,B476,118;%%

\bibitem{CT}
  M.~Cveti\v c and A.A.~Tseytlin,
 {\it Solitonic strings and BPS saturated dyonic black holes,}
  Phys.\ Rev.\ {\bf D53}, 5619 (1996)
  [Erratum-ibid.\ {\bf D55}, 3907 (1997)],
  hep-th/9512031.
  %%CITATION = HEP-TH/9512031;%%

  \bibitem{CYIII}
  M.~Cveti\v c and D.~Youm,
 {\it All the static spherically symmetric black holes of heterotic 
string on a six torus,}
  Nucl.\ Phys.\ {\bf B472}, 249 (1996),
 hep-th/9512127.
  %%CITATION = HEP-TH/9512127;%%

  \bibitem{L}
  F. Larsen,
{\it A string model of black hole microstates},
Phys. Rev. {\bf D56}, 1005 (1997), hep-th/9702153.
 %%CITATION = HEP-TH/9702153;%%

\bibitem{CLI}
M. Cveti\v c and F. Larsen,
{\it General rotating black holes in string theory: Greybody
factors and  event horizons},
Phys. Rev. {\bf D56}, 4994 (1997), hep-th/9705192.
  %%CITATION = HEP-TH/9705192;%%

\bibitem{CLII}
   M. Cveti\v c and F. Larsen,
{\it Greybody factors for rotating black holes in four dimensions},
Nucl. Phys. {\bf B506}, 107 (1997), hep-th/9706071.
 %%CITATION = HEP-TH/9706071;%%

 \bibitem{CLIII} M. Cveti\v c and F. Larsen,
{\it Greybody factors and charges in Kerr/CFT},
JHEP {\bf 0909}, 088 (2009), arXiv:0908.1136 [hep-th].

 \bibitem{Ansorg1}
  M. Ansorg, J. Hennig and C. Cederbaum,
{\it Universal properties of distorted Kerr-Newman black holes},
  Gen.\ Rel.\ Grav.\  {\bf 43}, 1205 (2011),
  arXiv:1005.3128 [gr-qc].
  %%CITATION = ARXIV:1005.3128;%%

\bibitem{Ansorg7}
  J.L. Jaramillo, N. Vasset and M. Ansorg,
{\it A numerical study of Penrose-like inequalities in a family of axially
 symmetric initial data},
  arXiv:0712.1741 [gr-qc].
  %%CITATION = ARXIV:0712.1741;%%

\bibitem{CR}
A.~Castro and M.J.~Rodriguez,
 {\it Universal properties and the first law of black hole inner mechanics,}
  Phys.\ Rev.\  {\bf D86}, 024008 (2012),
  arXiv:1204.1284 [hep-th].
  %%CITATION = ARXIV:1204.1284;%%

\bibitem{OrtinI} 
  P.~Galli, T.~Ortin, J.~Perz and C.~S.~Shahbazi,
  {\it ``Non-extremal black holes of N=2, d=4 supergravity,}
  JHEP {\bf 1107}, 041 (2011)
  [arXiv:1105.3311 [hep-th]].
  %%CITATION = ARXIV:1105.3311;%%

\bibitem{OrtinII} 
  P.~Meessen and T.~Ortin,
 {\it Non-Extremal Black Holes of N=2,d=5 Supergravity,}
  Phys.\ Lett.\ B {\bf 707}, 178 (2012)
  [arXiv:1107.5454 [hep-th]].
  %%CITATION = ARXIV:1107.5454;%%

\bibitem{OrtinIII} 
  P.~Meessen, T.~Ortin, J.~Perz and C.~S.~Shahbazi,
 {\it Black holes and black strings of N=2, d=5 supergravity in the H-FGK formalism,}
  JHEP {\bf 1209}, 001 (2012)
  [arXiv:1204.0507 [hep-th]].
  %%CITATION = ARXIV:1204.0507;%%

 \bibitem{CMS}
  A.~Castro, A.~Maloney and A.~Strominger,
 {\it Hidden conformal symmetry of the Kerr black hole,}
  Phys.\ Rev.\ {\bf D82}, 024008 (2010),
  arXiv:1004.0996 [hep-th].
  %%CITATION = ARXIV:1004.0996;%%

\bibitem{MS}
J.M.~Maldacena and A.~Strominger,
  ``Black hole grey body factors and d-brane spectroscopy,''
  Phys.\ Rev.\  {\bf D55}, 861 (1997),
  hep-th/9609026.
  %%CITATION = PHRVA,D55,861;%%


%\GuicaMU
\bibitem{GHSS}
  M.~Guica, T.~Hartman, W.~Song and A.~Strominger,
{\it The Kerr/CFT correspondence},
  arXiv:0809.4266 [hep-th].
  %%CITATION = ARXIV:0809.4266;%%\bibitem{CL11I}

\bibitem{CL11I}
  M.~Cveti\v c and F.~Larsen,
{\it Conformal symmetry for general black holes,}
  JHEP {\bf 1202}, 122 (2012),
  arXiv:1106.3341 [hep-th].
  %%CITATION = ARXIV:1106.3341;%%

\bibitem{CL11II}
M. Cveti\v c and F. Larsen,
{\it Conformal symmetry for black holes in four dimensions,}
 JHEP {\bf 1209}, 076 (2012),
  arXiv:1112.4846 [hep-th].
  %%CITATION = ARXIV:1112.4846;%

  \bibitem{CG}
  M.~Cveti\v  c and G.W.~Gibbons,
{\it Conformal symmetry of a black hole as a scaling limit: a black hole 
in an asymptotically conical box,}
  JHEP {\bf 1207}, 014 (2012),
  arXiv:1201.0601 [hep-th].
  %%CITATION = ARXIV:1201.0601;%%

  \bibitem{Virmani}
A.~Virmani,
{\it Subtracted geometry from Harrison transformations,}
 JHEP {\bf 1207}, 086 (2012),
arXiv:1203.5088 [hep-th].
  %%CITATION = ARXIV:1203.5088;%%


\bibitem{CGZ}
  M.~Cveti\v c, M.~Guica and Z.H.~Saleem,
{\it General black holes, untwisted,}
  arXiv:1302.7032 [hep-th].
  %%CITATION = ARXIV:1302.7032;%%

\bibitem{VirmaniII}
  A.~Sahay and A.~Virmani,
{\it Subtracted geometry from Harrison transformations: II,}
  arXiv:1305.2800 [hep-th].
  %%CITATION = ARXIV:1305.2800;%%

\bibitem{KI}
  A.~Chakraborty and C.~Krishnan,
  {\it Subttractors,}
  arXiv:1212.1875 [hep-th].
  %%CITATION = ARXIV:1212.1875;%%


\bibitem{KII}
  A.~Chakraborty and C.~Krishnan,
  {\it Attraction, with boundaries,}
  arXiv:1212.6919 [hep-th].
  %%CITATION = ARXIV:1212.6919;%%

\bibitem{KIII}
  S.~Jana and C.~Krishnan,
  {\it A Kaluza-Klein subttractor,}
  arXiv:1303.3097 [hep-th].
  %%CITATION = ARXIV:1303.3097;%%

  \bibitem{deBoer}
  M.~Baggio, J.~de Boer, J.I.~Jottar and D.R.~Mayerson,
{\it Conformal symmetry for black holes in four dimensions and 
irrelevant deformations,}
  arXiv:1210.7695 [hep-th].
  %%CITATION = ARXIV:1210.7695;%%

\bibitem{cvgipo} M. Cveti\v c, G.W. Gibbons and C.N. Pope,
{\it Universal area product formulae for rotating and charged black holes
in four and higher dimensions},
Phys.\ Rev.\ Lett.\  {\bf 106}, 121301 (2011),
arXiv:1011.0008 [hep-th].
%%CITATION = ARXIV:1011.0008;%%

\bibitem{CDGK}
  A.~Castro, N.~Dehmami, G.~Giribet and D.~Kastor,
 {\it On the universality of inner black hole mechanics and higher 
curvature gravity,}
  arXiv:1304.1696 [hep-th].
  %%CITATION = ARXIV:1304.1696;%%

\bibitem{liulu} H.-S. Liu and H.~L\"u,
{\it Charged rotating AdS black hole and its thermodynamics in conformal
gravity},
JHEP {\bf 1302}, 139 (2013), arXiv:1212.6264 [hep-th].
%%CITATION = ARXIV:1212.6264;%%

\bibitem{cclp} Z.-W. Chong, M. Cveti\v c, H. L\"u and C.~N. Pope,
{\it Charged rotating black holes in four-dimensional gauged and ungauged
supergravities},
Nucl.\ Phys.\ {\bf B717}, 246 (2005), hep-th/0411045.
%%CITATION = HEP-TH/0411045;%%

\bibitem{duff} L. Borsten, D. Dahanayake, M.J. Duff, H. Ebrahim and W. Rubens,
{\it Black holes, qubits and octonions},
  Phys.\ Rept.\  {\bf 471}, 113 (2009),  arXiv:0809.4685 [hep-th].
%%CITATION = ARXIV:0809.4685;%%

\bibitem{CH}
M.~Cveti\v c and C.M.~Hull,
 {\it Black holes and U duality},
  Nucl.\ Phys.\ {\bf B480}, 296 (1996), hep-th/9606193.
  %%CITATION = NUPHA,B480,296;%%

\bibitem{wu} S.-Q. Wu,
{\it General nonextremal rotating charged AdS black holes in five-dimensional
$U(1)^3$ gauged supergravity: A simple construction method},
  Phys.\ Lett.\ {\bf B707}, 286 (2012),
 arXiv:1108.4159 [hep-th].
%%CITATION = ARXIV:1108.4159;%%

\bibitem{Rasheed}
  D.~Rasheed,
 {\it The rotating dyonic black holes of Kaluza-Klein theory,}
  Nucl.\ Phys.\ {\bf B454}, 379 (1995),
hep-th/9505038.
  %%CITATION = HEP-TH/9505038;%%

  \bibitem{LarsenKK}
  F.~Larsen,
 {\it Rotating Kaluza-Klein black holes,}
  Nucl.\ Phys.\ {\bf B575}, 211 (2000),
hep-th/9909102.
  %%CITATION = HEP-TH/9909102;%%


%\cite{Moon:2011hq}
\bibitem{Moon:2011hq}
  T.~Moon, Y.S.~Myung and E.J.~Son,
{\it $f(R)$ black holes,}
  Gen.\ Rel.\ Grav.\  {\bf 43}, 3079 (2011),
  arXiv:1101.1153 [gr-qc].
  %%CITATION = ARXIV:1101.1153;%%
  %21 citations counted in INSPIRE as of 02 Jun 2013

%\cite{Larranaga:2011fv}
\bibitem{Larranaga:2011fv}
  A.~Larranaga,
{\it A rotating charged black hole solution in $f(R)$ gravity,}
  Pramana {\bf 78}, 697 (2012),
  arXiv:1108.6325 [gr-qc].
  %%CITATION = ARXIV:1108.6325;%%
  %6 citations counted in INSPIRE as of 02 Jun 2013

%\cite{Riegert:1984zz}
\bibitem{Riegert:1984zz}
  R.J.~Riegert,
{\it Birkhoff's theorem in conformal gravity,}
  Phys.\ Rev.\ Lett.\  {\bf 53}, 315 (1984).
  %%CITATION = PRLTA,53,315;%%
  %47 citations counted in INSPIRE as of 02 Jun 2013

\bibitem{BC}
T. Birkandan and M. Cveti\v c, unpublished.

%\cite{HL}
\bibitem{HL}
  H.~L\"u, {\it Charged dilatonic AdS black holes and magnetic 
AdS$_{D-2} \times  R^2$ vacua,}
  arXiv:1306.2386 [hep-th].
  %%CITATION = ARXIV:1306.2386;%%
\end{thebibliography}
\end{document}